\documentclass[12pt]{article}
\usepackage{graphicx}
\usepackage{here}
\def\beq{\begin{equation}}
\def\eeq{\end{equation}}
\def\bea{\begin{eqnarray}}
\def\eea{\end{eqnarray}}
\def\bq{\begin{quote}}
\def\eq{\end{quote}}

\def\bl{\bullet}
\def\nnb{\nonumber}
\def\ga{\left(}
\def\dr{\right)}

\def\lb{\lbrack}
\def\rb{\rbrack}

\def\rar{\rightarrow}
\def\nnb{\nonumber}
\def\la{\langle}
\def\ra{\rangle}
\def\nin{\noindent}
\def\ba{\begin{array}}
\def\ea{\end{array}}
\def\bm{\overline{m}}
\begin{document}
\topmargin -1.5cm
\oddsidemargin -.5cm
\evensidemargin -1.0cm
\pagestyle{empty}
\begin{flushright}
\end{flushright}
\vspace*{1cm}
\begin{center}
\section*{Strange Quark, Tachyonic Gluon Masses \\
and \boldmath $|V_{us}|$ from Hadronic Tau Decays }
\vspace*{0.5cm}
{\bf Stephan Narison} \\
\vspace{0.3cm}
Laboratoire de Physique Th\'eorique et Astrophysiques\\
Universit\'e de Montpellier II\\
Place Eug\`ene Bataillon\\
34095 - Montpellier Cedex 05, France\\
\vspace*{1.5cm}
{\bf Abstract} \\ \end{center}
\vspace*{2mm}
\noindent
We apply, to the new OPAL data on $V$+$A$ strange
spectral functions, a suitable combination of QCD spectral sum rules directly sensitive to the combination
of strange quark ($m_s$) and tachyonic gluon ($\lambda^2$) masses . Using the mean value of $\lambda^2$ from
a sum rule analysis of different channels, we deduce the invariant strange
quark mass $\hat m_s= (106^{+33}_{-37})$ MeV to order $\alpha_s^3$, which leads to the
running mass
$\overline m_s$(2 GeV)= ($93^{+29}_{-32}$) MeV. We also obtain an
interesting though less accurate estimate of the CKM angle: 
$|V_{us}|=0.217\pm 0.019$.
\vspace*{3.0cm}
\begin{flushleft}
\end{flushleft}
\vfill\eject
\setcounter{page}{1}
 \pagestyle{plain}
\section{Introduction} \par
The determination of the strange quark running mass  is of prime
importance for low-energy phenomenology, for CP-violation
and for SUSY-GUT or some other model-buildings.
Since
the advent of QCD, where a precise meaning for
the definition of the running quark masses within 
the $\overline{MS}$-scheme \cite{FLO} has been provided,
a large number of
efforts have been devoted to the determinations of the strange quark mass \footnote{For reviews, see e.g.
\cite{GASSER,SNL,SNB}.}
using QCD spectral sum rules  \footnote{For a review see e.g. \cite{SNB}.} \`a la
SVZ \cite{SVZ}, in the pseudoscalar
\cite{PSEUDO}, the scalar
\cite{SCAL}, the $e^+e^-$ \cite{REIND,SNE} channels, 
tau-decay data \cite{TAU,PICH,KUHN} and lattice simulations
\cite{LATT}, while some bounds have been also derived from the positivity of the spectral functions 
and from the extraction of the quark condensate \cite{BOUND,SNL}.
\\
\nin
In the following, we propose a
new method for determining the combination of the strange quark mass and tachyonic gluon mass ($\lambda^2$), which will allow us to
test the effect of this new $\lambda^2$-term not present in the standard OPE.
This $\lambda^2$-term has been introduced in \cite{ZAK} in order to mimic the UV renormalon
contribution in the resummation of the PT series, where it is expected to replace the uncalculated infinite number of terms of the PT series.
Its value has been estimated from
$e^+e^-$ into hadrons data
\cite{CNZ,SNV} to be:
${(\alpha_s/\pi)}\lambda^2\simeq -(0.06\pm 0.03)~{\rm GeV}^2~$, while  the pseudoscalar channel and a
fit of the lattice data in the $x$-space for the (pseudo)scalar and V+A channels leads to ${(\alpha_s/\pi)}\lambda^2\simeq -(0.12\pm
0.06)~{\rm GeV}^2~$. In the following, we shall consider the average of the previous estimates:
\beq\label{eq: lambda}
d_2\equiv {(\alpha_s/\pi)}\lambda^2\simeq -(0.07\pm 0.03)~{\rm GeV}^2~,
\eeq 
which is almost scale independent.
One should notice that, the effect of this term which is relatively negligible does not perturb and, in some cases, improves the
well-established existing sum rules results \cite{SNV,CNZ,CN} and the precise determination of $\alpha_s$ from $\tau$-decays \cite{CNZ,SN05}
obtained originally in
\cite{BNP}. It also solves
\cite{CNZ} the sum rule puzzle scales noticed by \cite{NSVZ} in the pseudoscalar pion and gluonia channels, where the scales are much larger than the one
of the
$\rho$ meson. We shall further study the effect of
$d_2$ in the determination  of the strange quark mass from tau decays. In previous analysis of this channel
\cite{TAU,PICH,KUHN}, inspired from the first analysis of tau-decay data \cite{BNP}, the
$\lambda^2$-effect, which is flavour independent, is absent, to leading order of PT, as the authors work with the difference of the  spectral functions
in the $\bar ud$ and $\bar us$ channels (so-called flavour breaking sum  rules). However, the price to pay is the large
cancellation between the two different spectral functions, implying a large error bar in the final result. Some other eventual
problems appearing in these analysis, are the decrease of the value of the strange quark mass output  (absence of stability in the number of
moments) and the deterioration of the convergence of the OPE for increasing dimension of the moments, and the
non-trivial separation of the spin zero and one parts of the spectral functions in some of the analysis.  In this paper, we avoid these problems by
working with a given flavoured
$\bar us$ spectral function involving the sum of the  spin zero and one mesons, but in the same time, our analysis will be affected by $\lambda^2$.
This spectral function has been measured by ALEPH, OPAL and CLEO
\cite{EXP},and more recently by OPAL \cite{OPAL04}. 
\section{The sum rules and the QCD expression}
We shall be concerned with the two-point correlator:
\beq
\Pi^{V+A}_{\mu\nu}(q^2)\equiv i\int d^4x~e^{iqx}\la 0|{\cal T}J_\mu(x)
J^\dagger_\nu(0)
|0\ra
=\ga q_\mu q_\nu-q^2g_{\mu\nu}\dr \Pi_{V+A}^{(1+0)}(q^2)+q^2g_{\mu\nu} \Pi_{V+A}^{(0)}(q^2)
\eeq
built from the charged local weak current:
\beq
J_\mu=\bar u\gamma_\mu(1-\gamma_5)s~.
\eeq
The upper indices 0, 1 refer to the corresponding spin of hadrons entering into the spectral function. Following SVZ
\cite{SVZ}, the correlator can be approximated by:
\beq
\Pi_{V+A}(Q^2)\simeq\sum_{d\geq 0}{{\cal O}_{2d}\over (Q^2)^d}
\eeq
where ${\cal O}_{2d}\equiv C_{2d}\la{\cal O}_{2d}\ra$ is the short hand-notation of the QCD
non-perturbative condensates $\la{\cal O}_{2d}\ra$ of dimension
$D\equiv 2d$ and its associated perturbative Wilson coefficient $C_{2d}$; 
$q^2\equiv -(Q^2 >0)$ is the momentum transfer. 
The spectral function $(v+a)$:
\beq
\frac{1}{\pi}{\rm Im}\Pi_{V+A}\equiv\frac{1}{2\pi^2}\ga v+a\dr~.
\eeq
has been measured using
$\tau$-decay data, via:
\bea
v_1/a_1&=& {M^2_\tau\over 6|V_{us}|^2S_{ew}}
\ga 1-{t\over M_\tau^2}\dr^{-2}\ga 1+{2t\over M_\tau^2}\dr^{-1}
{{B\ga\tau\rar
(V/A)^{\ga S=-1,~J=1\dr} +\nu_\tau\dr}\over {B\ga\tau\rar
e^-\bar \nu_e+\nu_\tau\dr}}
\frac{1}{ N_{V/A}}{dN_{V/A}\over dt}\nnb\\
v_0/a_0&=& {M^2_\tau\over 6|V_{us}|^2S_{ew}}
\ga 1-{t\over M_\tau^2}\dr^{-2}
{{B\ga\tau\rar
(V/A)^{\ga S=-1,~J=0\dr} +\nu_\tau\dr}\over {B\ga\tau\rar
e^-\bar \nu_e+\nu_\tau\dr}}
\frac{1}{ N_{V/A}}{dN_{V/A}\over dt}~,
\eea
where OPAL has used $|V_{us}|=0.2196\pm 0.0023$ \cite{PDG}, $M_\tau=1776.9^{+0.31}_{-0.27}$ MeV
\cite{BES} and
$S_{ew}=1.0194\pm 0.0040$ \cite{MARC}.\\
\nin
For a pedagogical purpose, we write the QCD expression of the $V$+$A$ correlator to leading
order in $\alpha_s$ and $m_s$(see e.g.
\cite{BNP,SNB}), and including the leading new $\lambda^2$-tachyonic gluon term \cite{CNZ}:
\bea\label{sum}
\Pi^{(0+1)}_{V+A}(Q^2)={1\over
2\pi^2}\Bigg{\{}-
\ln {Q^2\over\nu^2}-
{{\cal O}_2^0\over Q^2} +
{{\cal O}_4^0\over Q^4}
+{{\cal O}_6^0\over
Q^6}\Bigg{\}}~,
\eea
with obvious notations, where:
\bea
{\cal O}_2^0&=&{3}m_s^2+{\alpha_s\over
\pi}{\lambda^2}\nnb\\
{\cal O}_4^0&=& 4\pi^2\ga m_s\la\bar ss\ra+m_u\la\bar uu\ra\dr+{\pi\over
3}{\la\alpha_s(G^a_{\mu\nu})^2\ra}\nnb\\
{\cal O}_6^0&=&{256\pi^3\over
81}{\alpha_s\la\bar{u}u\ra^2}
\eea
Its inverse Laplace transform sum rule (LSR) reads
\cite{SVZ,NR,BELL,SNB}:
\beq
{\cal L}_{0}\equiv\int_0^{t_c} dt ~e^{-t\tau}\frac{1}{\pi}{\rm Im}\Pi^{(0+1)}_{V+A} =
{\tau^{-1}\over
2\pi^2}\Bigg{\{} 1-e^{-t_c\tau}-{\cal O}^0_2{\tau}+{\cal O}^0_4\tau^2+{1\over 2}{\cal O}_6^0\tau^3\Bigg{\}}~,
\eeq 
from which, one
can derive, to leading order in $\alpha_s$ and $m_s$, the LSR:
\beq
{\cal L}_1\equiv -{d\over d\tau}{\cal L}_0
\equiv \int_0^{t_c} dt ~te^{-t\tau}\frac{1}{\pi}{\rm Im}\Pi^{(0+1)}_{V+A} =
{\tau^{-2}\over
2\pi^2}\Bigg{\{} 1-(1+t_c\tau)e^{-t_c\tau}-{\cal O}^0_4\tau^2-{\cal O}_6^0\tau^3\Bigg{\}}~,
\eeq 
and the leading order FESR
 \cite{BERTL,SNB}:
\bea\label{eq: fesr}
{\cal M}_{0}&\equiv&\int_0^{t_c} dt ~\frac{1}{\pi}{\rm Im}\Pi^{(0+1)}_{V+A} =
{t_c\over
2\pi^2}\Bigg{\{} 1-{{\cal O}^0_2\over t_c}\Bigg{\}},
\nnb\\ {\cal M}_{1}&\equiv&\int_0^{t_c} dt
~t~\frac{1}{\pi}{\rm Im}\Pi^{(0+1)}_{V+A} = 
{t_c^2\over
4\pi^2}\Bigg{\{} 1-
2{{\cal O}^0_4\over t_c^2}\Bigg{\}}~.
\eea
From these previous sum rules, one can derive the combination of sum rules \footnote{A sum rule similar to ${\cal N}_{10}$
has been used for the first time in the pseudoscalar channel for testing the size of the $SU(3)$ breakings in the
kaon PCAC relation \cite{SNK}.}:
\bea\label{eq:combine}
{\cal N}_{10}&\equiv&{\cal L}_0-\tau {\cal L}_1\equiv \int_0^{t_c} dt \ga 1-{t\tau}\dr e^{-t\tau}
\frac{1}{\pi}{\rm Im}\Pi^{(0+1)}_{V+A} \nnb\\
{\cal S}_{10}&\equiv& {\cal M}_{0}-{2\over t_c}{\cal M}_{1}\equiv \int_0^{t_c} dt \ga 1-2{t\over t_c}\dr
\frac{1}{\pi}{\rm Im}\Pi^{(0+1)}_{V+A}~,
\eea
which are sensitive, to leading order, to $m^2_s$ and $\lambda^2$. Unlike the individual sum rules, these
combinations of sum rules are less sensitive to the high-energy tail of the spectral functions (effect of the $t_c$
threshold). Then, one may expect that they are more accurate than the former. However, this accuracy may not be
comparable with the one of the $\tau$-decay like-sum rules where threshold effect suppresses completely the
effects near the real axis. \\ We shall also work with the ratio of moments:
\bea\label{eq: ratio}
{\cal R}_{10}\equiv 2{{\cal M}_{1}\over {\cal M}_{0}}~, 
\eea
which will also be useful for testing the duality between
the LHS (experiment) and RHS (QCD theory).
\section{QCD corrections and RGI parameters}
In order to account for the radiative corrections,
one introduces the expressions of the running coupling and masses. \\
$\bl$ To three-loop accuracy, the running
coupling can be parametrized as \cite{BNP,SNB}:
\bea
a_s(\nu)&=&a_s^{(0)}\Bigg\{ 1-a_s^{(0)}\frac{\beta_2}{\beta_1}\log
\log{\frac{\nu^2}{\Lambda^2}}\nnb \\
&+&\ga a_s^{(0)}\dr^2\lb\frac{\beta_2^2}{\beta_1^2}\log^2\
\log{\frac{\nu^2}{\Lambda^2}}-\frac{\beta_2^2}{\beta_1^2}\log
\log{\frac{\nu^2}{\Lambda^2}}-\frac{\beta_2^2}{\beta_1^2}
+\frac{\beta_3}{\beta_1}\rb+{\cal{O}}(a_s^3)\Bigg\},
\eea
with:
\beq
a_s^{(0)}\equiv \frac{1}{-\beta_1\log\ga\nu/\Lambda\dr}
\eeq
and
$\beta_i$ are the  ${\cal{O}}(a_s^i)$ coefficients of the
$\beta$-function in the $\overline{MS}$-scheme, which read for three flavours \cite{SNB}:
\beq
\beta_1=-9/2,~~~~\beta_2=-8,~~~~\beta_3=-20.1198.
\eeq
$\bl$ The expression of the running quark mass in terms of the
invariant mass $\hat{m}_i$ is \cite{FLO,SNB}:
\bea\label{eq: mshat}
&&\bm_i(\nu)=\hat{m}_i\ga -\beta_1 a_s(\nu)\dr^{-\gamma_1/\beta_1}
\Bigg\{1+\frac{\beta_2}{\beta_1}\ga \frac{\gamma_1}{\beta_1}-
 \frac{\gamma_2}{\beta_2}\dr a_s(\nu)\nnb \\
&&+\frac{1}{2}\Bigg{[}\frac{\beta_2^2}{\beta_1^2}\ga \frac{\gamma_1}
{\beta_1}-
 \frac{\gamma_2}{\beta_2}\dr^2-
\frac{\beta_2^2}{\beta_1^2}\ga \frac{\gamma_1}{\beta_1}-
 \frac{\gamma_2}{\beta_2}\dr+
\frac{\beta_3}{\beta_1}\ga \frac{\gamma_1}{\beta_1}-
 \frac{\gamma_3}{\beta_3}\dr\Bigg{]} a^2_s(\nu)+
1.95168a_s^3(\nu)\Bigg\},
\eea
where $\gamma_i$ are the ${\cal{O}}(a_s^i)$ coefficients of the
quark-mass anomalous dimension, which read for three flavours \cite{SNB}:
\beq
\gamma_1=2,~~~~\gamma_2=91/12,~~~~\gamma_3=24.8404.
\eeq
$\bl$ To order $\alpha_s^3$, the perturbative expression of the correlator reads, in terms of the running coupling evaluated
at $Q^2=\nu^2$ (see e.g. \cite{BNP,SNB}):
\beq
-Q^2{d\over dQ^2}\Pi_{V+A}(Q^2)= {1\over 2\pi^2}\Bigg{\{}1+\ga a_s\equiv 
\frac{\alpha_s(Q^2)}{\pi}\dr+1.6398
a_s^2+6.3711a_s^3+...\Bigg{\}},
\eeq
$\bl$ The $D=2$ contribution reads to order $\alpha_s^3$, in terms of the running mass and by including the  
$\lambda^2$ term \cite{BNP,CNZ,SNB,KUHN}:
\beq\label{eq:o2}
 Q^2\Pi_{V+A}^{(D=2)}(Q^2)\simeq -{1\over 2\pi^2}\Bigg{\{}a_s{\lambda^2}+{3}{\overline m_s^2}\ga
1+2.333a_s+19.58a_s^2+202.309a_s^3+ {\cal O}(a_s^4)\dr\Bigg{\}}~.
\eeq
The coefficient of the $a_s^4$ term has been estimated \cite{KUHN} using PT
optimisation schemes arguments to be $K\simeq 2276\pm 200$. In our approach its effect like
all unknown higher order terms will be mimiced by the $\lambda^2$-term present in the $D=4$ contribution
given below.\\
\nin
$\bl$ The $D=4$ contributions read \cite{BNP,SNB}:
\bea\label{eq: d4}
 Q^4\Pi_{V+A}^{(D=4)}(Q^2)&\simeq&{2}\Bigg{\{}
\frac{1}{12\pi}\ga 1-{11\over 18}a_s\dr{\la\alpha_s G^2\ra}\nnb\\
&+&\ga 1-a_s\dr\la m_u\bar uu+m_s\bar ss\ra+{4\over 27}a_s\la m_s\bar uu+m_u\bar ss\ra\nnb\\
&+&{1\over 4\pi^2}\ga -{12\over 7}a_s^{-1}+1\dr\bm_s^4-{1\over 28\pi^2}\Big{[}1-\ga {65\over
6}-16\zeta(3)\dr a_s\Big{]}\bm_s^4\nnb\\
&+&{1\over 4\pi^2}\ga -{25\over 3}+4\zeta(3)\dr m_s^2a_s\lambda^2
\Bigg{\}},\nnb\\
\eea
where the last term is due to the $\lambda^2$ term \cite{CNZ}, and $\zeta(3)$=1.202...
We shall use as input $\Lambda_3=(375\pm 25)$ MeV for three flavours, the value of $a_s\lambda^2$ in Eq. (\ref{eq:
lambda}) and
\cite{SNG,SNB,TARRACH}:
\bea
(m_u+m_d)\la \bar uu
+\bar dd\ra &=&-{2}m_\pi^2 f_\pi^2 \nnb \\
 (m_s+m_u)\la\bar ss+ \bar uu\ra&\simeq &
-{2}\times 0.7m_K^2 f_K^2\nnb\\
\la\alpha_s G^2\ra&\simeq&(0.07\pm 0.01)~\mbox{GeV}^4\nnb \\
\rho\alpha_s\la\bar uu\ra^2&\simeq& (5.8\pm 0.9)\times 10^{-4}~{\rm GeV}^6~,
\eea
where: $f_\pi=93.3$ MeV, $f_K=1.2 f_\pi$ and we have taken into account
a possible violation of kaon PCAC as suggested by the
QSSR analysis \cite{SNK,SNB}; $\rho\simeq 2-3$ is the measures of the deviation from the vacuum saturation estimate of the four-quark
condesates.\\
$\bl$ FESR including radiative corrections can be deduced from the previous expressions of the correlator
using either the Laplace or the Gaussian transforms \cite{BERTL,SNB}. For the FESR, the most dominant contributions
induced by the radiative corrections to the unit operator is the dimension-two terms:
\beq
\delta {\cal O}_2=-\ga{t_c\over 2\pi^2}\dr\ga{\beta_1\over 4}\dr a_s^2~,
\eeq
which adds to the contribution of the terms in Eq. (\ref{eq:o2}). One can check that this term remains a correction
of the $\lambda^2$ contribution in this previous equation. Some other corrections induced by the log-dependence of the
running terms taken into account in the numerical analysis remains also tiny corrections.
\section{Phenomenological analysis}
We parametrize the spectral function by using the most recent OPAL data discussed in a previous section
for
$t$ until
$M^2_\tau$. In so doing, we parametrize the data using standard polynomials fits which delimit the domain 
spanned by the error bars of the data. At the level of accuracy of about 30\%, where $m_s$ will be determined,
the inclusion of the correlations of each data points are not necessary. In fact, we have done similar methods
in parametrizing the $\tau$ and $e^+e^-$ data in the most accurate determination 
of the $g-2$ of the muon, where the error in the estimate is at the level of 1\% \cite{SNGM2}. 
In this accurate example, our result and the errors agreed quite well with ones
where the correlations among different data points have been taken into account \cite{YND}.\\
For  $t$ above
$M^2_\tau$, we add the QCD step function:
\beq
{1\over \pi}{\rm Im} \Pi_{V+A}(t\geq M^2_\tau)\simeq \theta (t-M^2_\tau){1\over 2\pi^2}
\Bigg{\{}1+a_s(t)+1.6398a_s^2(t)-10.284a_s^3(t)\Bigg{\}}~,
\eeq
which is consistent with the data at $t=M^2_\tau$.
\subsection*{Test of duality}
In principle, the value of the $t_c$-cut of the FESR integrals is a free parameter. We fix its optimal value by looking
for the region where the phenomenological and QCD sides of the ratio of moments in Eq. (\ref{eq: ratio}) are
equal. We present this analysis in Fig. 1, by showing the value of $t_c$ predicted by the sum rule versus $t_c$
and by comparing the result with the exact solution
$t_c=t_c$ for all values of
$t_c$. From Fig. 1, one
can see that the upper values of the data points provide stronger constraints on $t_c$ than their central value.
Considering this stronger constraint, we deduce, that QCD duality is best obtained at:
\beq
t_c\simeq {\rm M}_\tau^2~,
\eeq
\begin{figure}[H]
\begin{center}
\includegraphics[width=7cm]{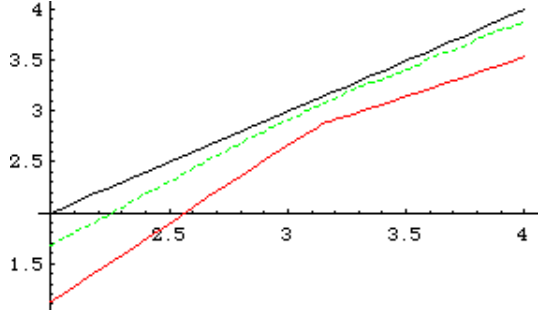}  
\caption{\footnotesize FESR prediction of $t_c$ versus $t_c$. The green curve corresponds the central value of the data;
the red one to the larger values of the data; the black continuous line is the exact solution $t_c=t_c$. The
curves correspond to the value $\hat m_s=(56-145)$ MeV and giving $a_s\lambda^2=-0.07$ GeV$^2$.}
\end{center}
\end{figure}
\nin
where one expects to get the optimal value of $m_s$ from FESR. In order to get this number, we have
used the value of the invariant mass
$\hat{m}_s=(56-145)$ MeV, which is a tiny correction in this duality test analysis.
Once we have fixed the value of $t_c$ where the best duality from the two sides of FESR has been obtained, we can now
estimate some other observables.
\subsection*{Estimate of \boldmath $\hat m_s$ versus \boldmath$\lambda^2$}
We can, in principle, estimate $\hat m_s$ using the combinations of LSR or/and FESR in Eq. (\ref{eq:combine}). A sample of
analysis is given in Fig. \ref{fig:estimate} for the LSR and in Fig. \ref{fig:estimateb} from the FESR:
\begin{figure}[H]
\begin{center}
\includegraphics[width=8cm]{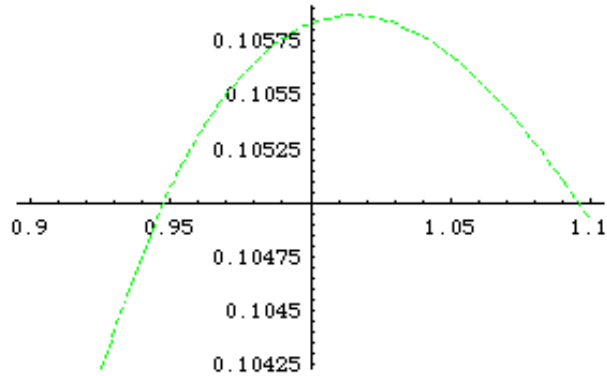}  
\caption{\footnotesize LSR prediction of $\hat m_s$ in GeV versus the LSR variable $\tau$ in GeV$^{-2}$ for
$-a_s\lambda^2=0.07$ GeV$^2$, using  the central value of the data and fixing $t_c=3.15$ GeV$^2$.}
\label{fig:estimate}
\end{center}
\end{figure}
\nin
\begin{figure}[H]
\begin{center}
\includegraphics[width=8cm]{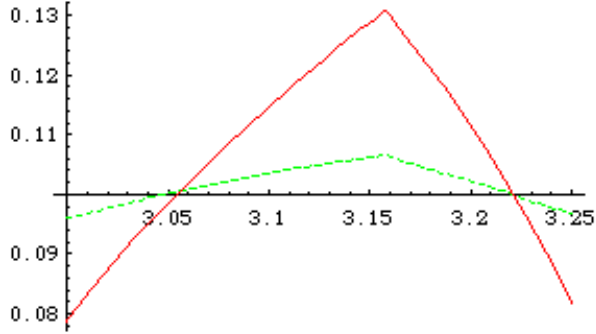}  
\caption{\footnotesize FESR prediction of $\hat m_s$ in GeV versus $t_c$ in GeV$^2$ for
$-a_s\lambda^2=0.07$ GeV$^2$. The green curve corresponds
to the central value of the data; the red one to the larger values of the data.}
\label{fig:estimateb}
\end{center}
\end{figure}
\nin
\begin{table*}[H]
\setlength{\tabcolsep}{1.5pc}
\begin{center}
\caption{Estimated value of $\hat m_s$ versus $\lambda^2$}
\label{tab: comparison}
\begin{tabular}[hbt]{lc}
&\\
\hline
\hline
\\
$-a_s\lambda^2$ in GeV$^2$&$\hat m_s$ in MeV \\
\\
\hline
\hline
\\
0.02& $53\pm 38\pm 8$\\
0.04& $79\pm 29\pm 8$\\
0.06&$98\pm 24\pm 9$\\
0.07&$106\pm 23\pm 9$\\
0.08&$114\pm 21\pm 9$\\
0.10&$128\pm 19\pm 11$\\
0.12&$140\pm 18\pm 11$\\
0.15&$157\pm 16\pm 11$\\
&\\
\hline
\hline
\end{tabular}
\end{center}
{\footnotesize 
\begin{quote}
The first error is due to the data, the second one to $\Lambda$. The error due to the choice of $t_c$ around the
duality region is quite small of about 3 MeV as can be seen in Fig. \ref{fig:estimateb}. 
\noindent
\end{quote}}
\end{table*}
\nin

\begin{itemize}
\item In Fig. \ref{fig:estimate}, we give the value of $\hat m_s$ versus
the LSR variable $\tau$ at given $t_c=M^2_\tau$. One can notice that, though the ``optimal" estimate looks
reasonnable, it is obtained at the value of $\tau$ of about 1 GeV$^{-2}$, where, at this scale, the PT series
of the $m_s^2$ coefficient behaves badly rendering this result inaccurate. Therefore, we do not consider this
result in our final estimate. One should notice that the size of this optimization scale is typical  for the LSR,
as the exponential amplifies the low energy contribution to the spectral function. This is not the case of the
FESR which acts in the opposite region of the spectral function.
\item In Fig. \ref{fig:estimateb}, we give the value of $\hat m_s$ from FESR versus the $t_c$ variable. Optimal
estimate is obtained for the $t_c$-value where the duality between the two sides of FESR is the best. One can
notice that unlike the case of LSR, FESR results are obtained at $t_c=3.15$ GeV$^2$, where the PT series in the
$m_s^2$ make sense. 
\item Results for different values of
$\lambda^2$ in the range given in Eq. (\ref{eq: lambda}) are shown in Table \ref{tab: comparison}. 
Considering the mean value of $\lambda^2$ from Eq. (\ref{eq:
lambda}), we deduce from Table~\ref{tab: comparison}, the estimate:
\beq
\hat m_s\simeq (106^{+33}_{-37})~{\rm MeV} ~~~{\rm
for}~~~~a_s\lambda^2\simeq -(0.07\pm 0.03)~{\rm GeV}^2~.
\eeq
Using the relation in Eq. (\ref{eq: mshat}):
\beq
\bm_s(2~{\rm GeV})\simeq 0.876~ \hat m_s~,
\eeq
we translate the result on $\hat m_s$ into the
value of the running mass at 2 GeV to order $\alpha_s^3$:
\bea\label{eq: msfinal}
\bm_s(2~{\rm GeV})&\simeq& (93_{-32}^{+29})~{\rm MeV}~.
\eea
\item This result can be compared with the recent determinations
from flavour breaking-sum rule in tau-decays \cite{PICH,KUHN}, which is not affected to leading
order by $\lambda^2$:
\beq
\bm_s(2~{\rm GeV})\simeq (81\pm 22)~{\rm MeV}~,
\eeq 
and with
the one: 
\beq
 \bm_s(2~{\rm GeV})\simeq (105\pm 26)~{\rm MeV}~,
\eeq
deduced from the pion sum rule to order $\alpha_s^3$, where the $\lambda^2$ term has been added \cite{CNZ,SNL,SNB}:
\beq
(\bm_u+\bm_d)(2~{\rm GeV})\simeq (8.6\pm 2.1)~{\rm MeV}~, 
\eeq
 plus the ChPT ratio $(m_u+m_d)/2m_s=24.4\pm 1.5$ \cite{GASSER}. 
From the previous values, we can deduce the average\footnote{We have not included in the average the value of $m_s$ from the vector current
in \cite{SNMS}, which is now under reconsideration.}:
\beq
 \la\bm_s(2~{\rm GeV})\ra\simeq (92\pm 15)~{\rm MeV}~,
\eeq
which is comparable with lattice determinations.
\item Finally, one can compare this result with the lower bounds obtained
from the positivity of spectral functions proposed in \cite{BOUND} and updated to order $\alpha_s^3$ to be $(71.4\pm 3.7)$ MeV in \cite{SNB}
(the inclusion of $\lambda^2$ decreases this value by 5\%), and from an independent lower bound of about 80-90 MeV derived in \cite{BOUND2}
from direct extractions of the quark condensate. Our result in Eq. \ref{eq: msfinal} is compatible with these bounds.
However, values of $m_s$  corresponding to $-\lambda^2\leq 0.04$ GeV$^2$ are less favoured by these bounds and by the previous estimates from other
channels.
\end{itemize}
\subsection*{Estimate of \boldmath $\vert V_{us}\vert$}
We can estimate the CKM angle by working with the FESR ${\cal M}_0$ or/and the LSR ${\cal L}_0$. 
FESR gives the result:
\beq\label{eq: v_us1}
|V_{us}|\simeq 0.215\pm 0.017~,
\eeq
while the LSR gives:
\beq\label{eq: v_us2}
|V_{us}|\simeq 0.219\pm 0.021~.
\eeq
One can notice that the result is not sensitive to the change of $m_s$ in the
range 56 to 145 MeV, which, a posteriori justifies the use of the exponential
Laplace sum rule (LSR), though, in the optimal region, the PT
series of the $m_s^2$ contribution behaves badly. 
We take, as a final result, the arithmetic average of the FESR and LSR results:
\beq\label{eq: v_usfinal}
|V_{us}|\simeq 0.217\pm 0.019~,
\eeq
which, despite the large error, is an interesting output per se as it shows the consistency of the
approach used to get $m_s$. 
\section{Conclusions}
Our result for the strange quark mass, in
Eq. (\ref{eq: msfinal}), from the strange spectral function of hadronic tau decays shows that
the one for  non-zero value of the tachyonic gluon mass is in better agreement with the existing
determinations of this quantity from $\tau$-decays and pseudoscalar channels and with the lower bounds derived from different sum rules
\cite{BOUND,BOUND2,SNL,SNB} than the one in the theory without a tachyonic gluon. \\ Our result in Eq. (\ref{eq: v_usfinal}) for $V_{us}$
is less accurate than existing determinations but interesting per se for testing the  consistency of the whole approach.\\
Improvments of the results obtained in this paper require more accurate data, which will also help for a much better determination of $m_s$
or/and of the tachyonic gluon mass $\lambda^2$.
\section*{Acknowledgements}
It is a pleasure to thank Valya Zakharov for multiple email communications and discussions.

\end{document}